\documentclass[conference]{IEEEtran}
\IEEEoverridecommandlockouts

\usepackage{cite}
\usepackage{graphicx}
\usepackage{url}
\usepackage{amsmath,amssymb,amsfonts}
\usepackage{algorithmic}
\usepackage{textcomp}
\usepackage{xcolor}
\usepackage{orcidlink}
\usepackage{threeparttable}
\usepackage{tabularx}
\usepackage{tikz}
\usetikzlibrary{arrows.meta,positioning,shapes.multipart}
\def\BibTeX{{\rm B\kern-.05em{\sc i\kern-.025em b}\kern-.08em
    T\kern-.1667em\lower.7ex\hbox{E}\kern-.125emX}}

\begin{document}

\title{Real-Time Emergency Vehicle Siren Detection with Efficient CNNs on Embedded Hardware
\thanks{This work was partially supported by the European Union under the Italian National Recovery and Resilience Plan (NRRP) of NextGenerationEU, partnership on “Telecommunications of the Future” (PE00000001 - program “RESTART”).}
}

\author{
\IEEEauthorblockN{1\textsuperscript{st} Marco Giordano \orcidlink{0009-0001-1649-6085}}
\IEEEauthorblockA{{\textit{\scriptsize{Dpt. of Information Engineering, Computer Science and Mathematics (DISIM)}}} \\
\textit{University of L'Aquila}\\
L'Aquila, Italy \\
marco.giordano3@graduate.univaq.it}
\and
\IEEEauthorblockN{2\textsuperscript{nd} Stefano Giacomelli \orcidlink{0009-0009-0438-1748}}
\IEEEauthorblockA{{\textit{\scriptsize{Dpt. of Information Engineering, Computer Science and Mathematics (DISIM)}}} \\
\textit{University of L'Aquila}\\
L'Aquila, Italy \\
stefano.giacomelli@graduate.univaq.it}
\and
\IEEEauthorblockN{3\textsuperscript{rd} Claudia Rinaldi \orcidlink{0000-0002-1356-8151}}
\IEEEauthorblockA{{\textit{\scriptsize{National Inter-University Consortium for Telecommunications (CNIT)}}} \\
\textit{University of L'Aquila}\\
L'Aquila, Italy \\
claudia.rinaldi@univaq.it}
\and
\IEEEauthorblockN{4\textsuperscript{th} Fabio Graziosi \orcidlink{0000-0001-7808-0707}}
\IEEEauthorblockA{{\textit{\scriptsize{Dpt. of Information Engineering, Computer Science and Mathematics (DISIM)}}} \\
\textit{University of L'Aquila}\\
L'Aquila, Italy \\
fabio.graziosi@univaq.it}
}

\maketitle

\begin{abstract}
We present a full-stack emergency vehicle (EV) siren detection system designed for real-time deployment on embedded hardware. The proposed approach is based on E2PANNs, a fine-tuned convolutional neural network derived from EPANNs, and optimized for binary sound event detection under urban acoustic conditions. A key contribution is the creation of curated and semantically structured datasets—AudioSet-EV, AudioSet-EV Augmented, and Unified-EV—developed using a custom AudioSet-Tools framework to overcome the low reliability of standard AudioSet annotations.

The system is deployed on a Raspberry Pi 5 equipped with a high-fidelity DAC+microphone board, implementing a multithreaded inference engine with adaptive frame sizing, probability smoothing, and a decision-state machine to control false positive activations. A remote WebSocket interface provides real-time monitoring and facilitates live demonstration capabilities.

Performance is evaluated using both framewise and event-based metrics across multiple configurations. Results show the system achieves low-latency detection with improved robustness under realistic audio conditions. This work demonstrates the feasibility of deploying IoS-compatible SED solutions that can form distributed acoustic monitoring networks, enabling collaborative emergency vehicle tracking across smart city infrastructures through WebSocket connectivity on low-cost edge devices.
\end{abstract}

\begin{IEEEkeywords}
Sound Event Detection, Emergency Vehicle Sirens, Raspberry Pi 5, Real-Time Inference, Embedded Systems, Urban Acoustics
\end{IEEEkeywords}

\section{Introduction}

Emergency vehicle (EV) siren detection is a key enabler of intelligent transportation systems, supporting real-time decision-making for autonomous vehicles, traffic monitoring infrastructures, and urban sound analysis platforms\cite{castorena_safety-oriented_2024,ramirez_siren_2022,dileep_review}. The ability to recognize sirens promptly from live audio streams enhances situational awareness, enabling applications such as automated braking, smart rerouting, and prioritization of emergency response.

Unlike vision-based approaches that rely on line-of-sight, audio-based emergency vehicle detection provides 360-degree awareness and operates effectively in occluded scenarios, at night, and in adverse weather conditions. This makes sound-based detection a critical complement to visual systems in autonomous navigation.

This work aligns with the emerging Internet of Sounds (IoS) paradigm, where distributed acoustic sensors form interconnected networks for collaborative environmental monitoring. Unlike isolated detection systems, our approach is designed to integrate seamlessly into IoS infrastructures, enabling city-wide emergency vehicle tracking through coordinated edge nodes. The WebSocket-based interface facilitates real-time data sharing between detection nodes, supporting applications such as traffic light preemption, route optimization for emergency responders, and acoustic-based situational awareness in smart cities.

Despite growing research interest in sound event detection (SED) for safety-critical contexts, current solutions face limitations in terms of real-time operability, generalization to urban noise conditions, and hardware deployability\cite{mittal_acoustic_2023}. Many deep learning (DL) approaches rely on heavy models or offline processing pipelines, while low-power embedded implementations often sacrifice accuracy and robustness\cite{miyazaki_ambulance_2013,meucci_real-time_2008}.

This paper addresses these challenges by introducing a real-time EV siren detection pipeline optimized for embedded platforms. Building upon Efficient/Emergency Pre-trained Audio Neural Networks (E2PANNs)\cite{giacomelli_large-scale_2025} - a specialization of EPANNs framework\cite{singh_e-panns_2023,kong_panns_2020} - we propose a lightweight, convolutional architecture tailored for real-time framewise inference of emergency siren events. The model is fine-tuned on a curated subset of AudioSet\cite{gemmeke_audio_2017} (AudioSet-EV)\cite{giacomelli_audioset-tools_2025}, and evaluated using both framewise and event-based metrics\cite{mesaros_sound_2021}. We introduce a set of post-inference processing strategies — including a moving average filter over the inference sequence, dynamic adjustment of frame width based on confidence thresholds, and a requirement for a minimum number of consecutive positive frames — designed to suppress spurious activations and minimize false positive detections while maintaining responsiveness and precision.

The system is deployed on a Raspberry Pi 5 equipped with a Raspiaudio Ultra++ DAC+mic board, supporting live audio input and efficient model inference. A companion remote interface provides runtime monitoring and performance feedback, enabling real-time visualization of classification results and inference metrics.

Our contributions are as follows:
\begin{itemize}
  \item We design and optimize a compact CNN-based siren detector based on E2PANNs\cite{giacomelli_large-scale_2025}, achieving real-time performance on Raspberry Pi 5.
  \item We release and utilize AudioSet-EV, a filtered and taxonomically harmonized subset of AudioSet specialized for EV detection.
  \item We propose and evaluate an adaptive framewise inference mechanism to improve latency-accuracy trade-offs and false positive rejection.
  \item We implement a full-stack embedded solution with remote access interface for real-time feedback and demo reproducibility.
\end{itemize}

The Internet of Sounds vision requires edge devices capable of both local intelligence and network collaboration. Our system addresses this dual requirement by combining efficient on-device inference with standardized IoT protocols. This enables scenarios such as:
\begin{itemize}
    \item Mesh networks of acoustic sensors providing redundant coverage at critical intersections.
    \item Federated learning where edge nodes collaboratively improve detection models without raw audio sharing.
    \item Integration with existing urban IoT infrastructure for holistic emergency response coordination.
\end{itemize}

To the best of our knowledge, this is the first publicly documented implementation of a CNN-based EV siren detector achieving real-time inference on embedded hardware with live audio input and adaptive detection logic.

\section{Related Work}

Emergency siren detection from audio has been the subject of increasing attention, with approaches ranging from handcrafted feature extraction and classical machine learning models to modern deep learning-based frameworks. While accuracy has improved significantly, especially under controlled conditions, a majority of these systems remain unsuitable for real-time deployment on embedded hardware due to high computational demands or lack of inference latency evaluation.

Early works, such as Castorena et al.~\cite{castorena_safety-oriented_2024}, proposed CRNN and YOLO-based models trained on synthetic mixtures of sirens and ambient noise, achieving high accuracy but requiring GPU-level resources for real-time inference. Similarly, Ram\'irez et al.~\cite{ramirez_siren_2022} utilized spectrograms and 2D CNNs to classify siren types in urban recordings, though their evaluation showed high sensitivity to real-world SNR fluctuations.

More efficient models have been proposed recently. Shams et al.~\cite{shams_acoustic_2024} combined EfficientNet with 1D CNNs and self-attention layers, demonstrating 209 ms inference time on short clips. However, their small dataset and lack of embedded testing limits generalizability. Mittal et al.~\cite{mittal_acoustic_2023} applied ensemble methods using CNNs, RNNs, and FC layers, reaching high classification accuracy at the cost of 1.5 s per inference.

Only a few studies focused on embedded or resource-constrained implementations. Miyazaki et al.~\cite{miyazaki_ambulance_2013} and Meucci et al.~\cite{meucci_real-time_2008} explored DSP and microcontroller-based siren detection using FFT and pitch tracking, but without deep learning or large-scale evaluation. Beritelli et al.~\cite{beritelli_automatic_2006} used LPCs for analog inference on TI DSPs, though false positives under real-world noise remained an issue.

In terms of datasets, benchmarks such as ESC-50\cite{piczak_esc_2015}, UrbanSound8K\cite{fuentes_urban_2022}, and FSD50K\cite{fonseca_fsd50k_2022} have limited relevance due to taxonomic ambiguity and label sparsity. Specialized corpora like LSSiren and SireNNet~\cite{shah_sirennet-emergency_2023,lssiren, yadav_audio_2023} offer more structured siren detection setups but are often small or lack diversity. Even though affected by weak labeling, AudioSet remains the most comprehensive source, and our derived subset, AudioSet-EV\cite{giacomelli_audioset-tools_2025}, ensures taxonomic clarity and reproducibility.

Evaluation standards have evolved to include framewise and event-based metrics\cite{mesaros_sound_2021}, yet false positive suppression and noise robustness remain underexplored. Our work extends this line by integrating adaptive post-processing, a novel real-time embedded deployment, and a monitoring interface for live metric inspection.

\section{Baseline Model Architecture and Training}

The architecture adopted in this work builds upon the Efficient Pruned Audio Neural Networks (EPANNs) framework\cite{singh_e-panns_2023,kong_panns_2020}, a structured pruning variant derived from the CNN14 model within the PANNs family. EPANNs were selected as the foundation of our system due to their favourable trade-off between computational cost and baseline performance across general-purpose audio tagging tasks.

Our contribution begins with a thorough assessment of EPANNs in the context of Emergency Vehicle (EV) siren detection.

\subsection{Motivations for Fine-tuning and Dataset Curation}

\begin{figure}
    \centering
    \includegraphics[width=1.0\linewidth]{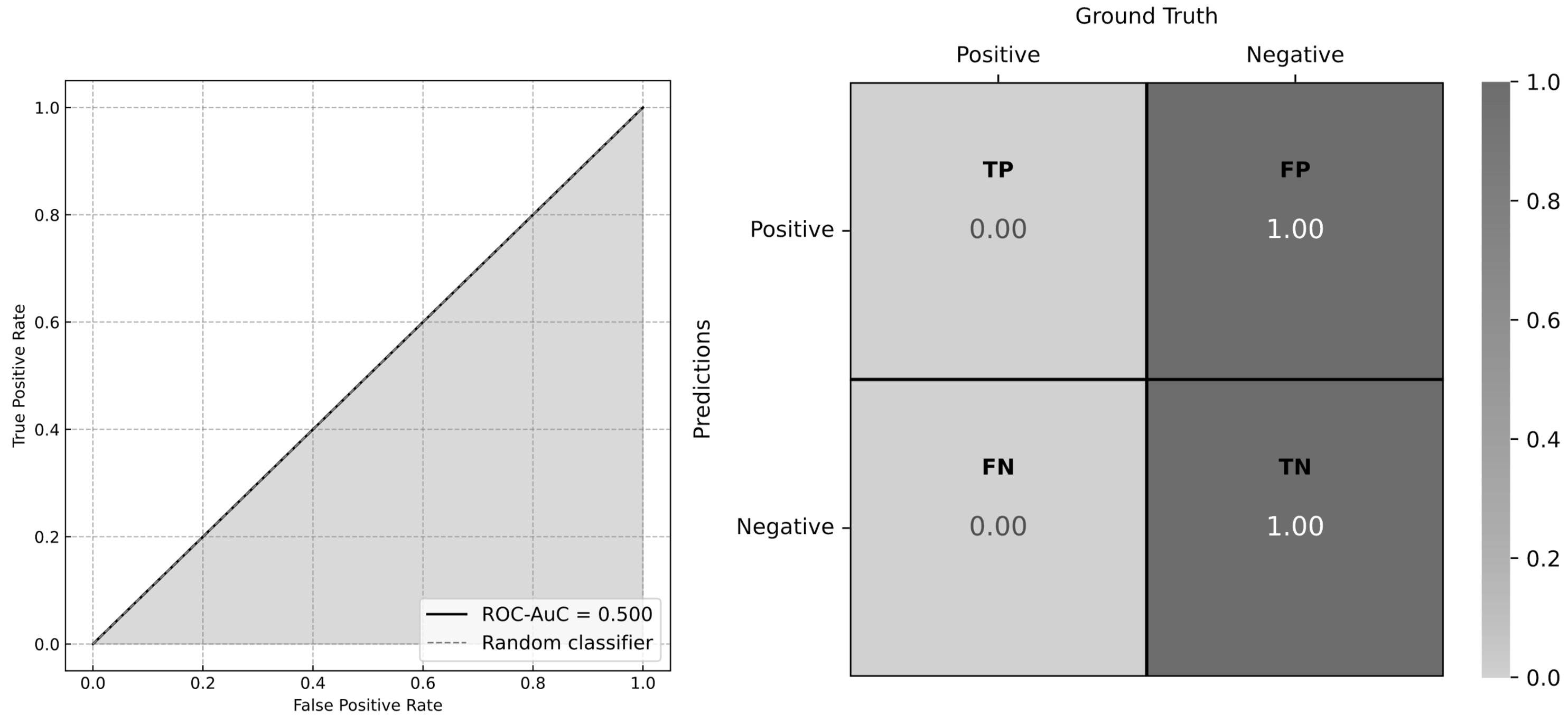}
    \caption{E-PANNs Confusion Matrix and ROC-AUC computed on preliminary AudioSet filtered samples.}
    \label{fig:epanns_pre_stats}
\end{figure}

As a preliminary step, we tested the pre-trained EPANNs model on an AudioSet-derived subset composed of one positive section (clips annotated with Emergency Vehicle label) and three negative sections (Traffic, Vehicles, and non-EV Alarms classes). This subset was designed to reflect a realistic and acoustically challenging deployment environment. Evaluation results revealed that EPANNs, despite its general capabilities, performed as a random classifier in this specific task configuration (see Fig. \ref{fig:epanns_pre_stats}) \cite{giacomelli_large-scale_2025}.

This observation prompted the need for a domain-adapted fine-tuning strategy and a more specialized dataset. To this end, we developed AudioSet-Tools\cite{giacomelli_audioset-tools_2025}, a Pytorch framework designed for precise label-based filtering, rebalancing, and reproducible dataset curation from AudioSet.

\subsection{AudioSet-EV and Dataset Variants} \label{datasets}

Using the AudioSet-Tools framework \cite{giacomelli_audioset-tools_2025}, we curated the following:

\begin{itemize}
    \item \textbf{AudioSet-EV:} a filtered subset consisting exclusively of EV siren and carefully selected negative samples, enforcing label disjointness\cite{giacomelli_audioset-tools_2025}.
    \item \textbf{AudioSet-EV Augmented:} an expanded version of the previous set with time-domain data augmentation applied online during training, including random noise injection, polarity inversion and temporal shifts.
    \item \textbf{Unified-EV:} a merged dataset combining the previous two with \textit{ESC-50}~\cite{piczak_esc_2015}, \textit{SireNNet}~\cite{shah_sirennet-emergency_2023}, \textit{LSSiren}~\cite{asif_large-scale_2022}, \textit{UrbanSound8K}~\cite{salamon_dataset_2014}, \textit{FSD50K}~\cite{fonseca_fsd50k_2022} for generalization testing.
\end{itemize}

These datasets underpin all subsequent training and evaluation phases.

\subsection{From EPANNs to E2PANNs} \label{subsec:e2panns}

We fine-tuned the EPANNs model on AudioSet-EV and AudioSet-EV Augmented using supervised training under a binary classification regime (EV vs. non-EV). Hyperparameter search experiments were conducted to identify optimal configurations for learning rate, dropout, and batch size\cite{giacomelli_large-scale_2025}.

Training inputs were 64-bin log-Mel spectrograms extracted from 10-second mono audio clips sampled at 32 kHz. Data augmentation included background mixing, time masking, and volume jitter.

Model evaluation was carried out using stratified training/validation splits of Unified-Ev dataset. The fine-tuned model, Efficient/Emergency Pre-trained Audio Neural Networks (E2PANNs), was exported as three best-performing checkpoints for subsequent validation on edge platform:

\begin{itemize}
    \item \texttt{Baseline\_EV}: top performing checkpoint trained on Audioset-EV without augmentations and with exponentially decaying learning rate.
    \item \texttt{Augmented\_EV}: top performing checkpoint trained on Audioset-EV with augmentations and fixed learning rate.
    \item \texttt{Transfer\_learning}: top performing checkpoint trained on the Unified-EV dataset \ref{datasets}.
\end{itemize}

\section{Real-Time Edge Computing Implementation} \label{sec:edge-computing}

Before deploying our E2PANNs model on the experimental target platform (Raspberry Pi 5) for real-time EV-SED experiments, we evaluated its ability to perform inference on small input sizes. The goal was to reduce computational complexity by leveraging the finetuned EPANNs to perform periodic inference on short, variable-length segments of streaming audio data. This approach aims to avoid recurrent layers — common in SoA SED systems — which introduce substantial computational and memory overhead.

To determine the minimum viable input size, we implemented a binary search algorithm~\cite{DBLP:journals/corr/ChadhaMM14} to identify the smallest input tensor that produces a valid model output. The minimum valid input size was found to be 9919 samples (approximately 310ms at 32kHz).

\begin{figure}
    \centering
    \includegraphics[width=1.0\linewidth]{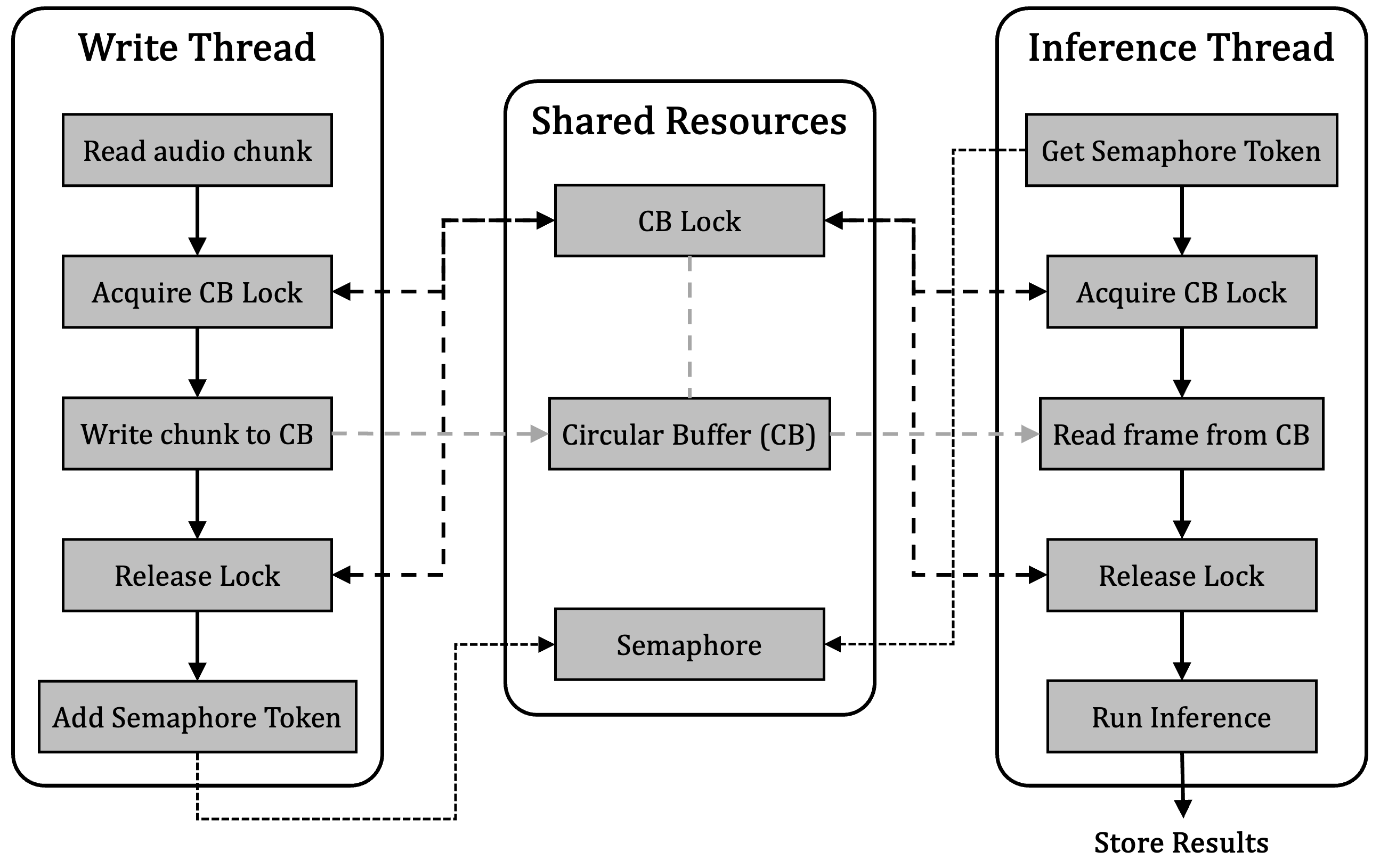}
    \caption{E2PANNs RT inference threading system running on Raspberry Pi 5.}
    \label{fig:RT_system}
\end{figure}

To support continuous EV siren detection on edge hardware, we designed a RT inference system tailored for the Raspberry Pi 5. The system (Figure~\ref{fig:RT_system}) adopts a multi-threaded architecture with explicit concurrency control, ensuring non-blocking and low-latency operation. Its core component is a \texttt{CircularBuffer} acting as shared memory between a \textit{producer} thread (writing audio data) and a \textit{consumer} thread (performing inference). Synchronization is enforced using Python’s \texttt{threading.Lock} and \texttt{threading.Semaphore} primitives.

The audio \texttt{Write Thread} (producer) emulates RT audio input by periodically writing short chunks from a source file or input buffer to the \texttt{CircularBuffer}. Thread-safe access is guaranteed by acquiring a \texttt{Lock} during buffer updates. Once data is written, a \texttt{Semaphore} token is released, signaling that new content is available for reading.

The \texttt{CircularBuffer} is a pre-allocated, fixed-size memory structure supporting wrapping reads and writes. It maintains a sliding window of the most recent samples, enabling continuous streaming behavior. Read/write operations are guarded by a \texttt{Lock}, while the \texttt{Semaphore} ensures reads occur only after new data has arrived.

The input \texttt{Frame Provider} module, embedded in the inference thread, manages frame extraction. Before reading, it acquires a \texttt{Semaphore} token (blocking if necessary), then locks the buffer to extract a valid audio frame. Frame length is dynamically adjusted based on the most recent model output: starting from the minimum valid size, the duration is increased whenever the output probability exceeds a predefined threshold. This adaptive mechanism progressively expands the temporal context during likely positive detections, enhancing robustness while limiting computational overhead when EV-like characteristics are absent. The frame duration is bounded by a configurable maximum.

The \texttt{Inference Thread} (consumer) runs concurrently, querying the \texttt{Frame Provider} for the next input and passing it to E2PANNs. The resulting probability is appended to a shared output list for decision logic. All inference times and timestamps are logged using a custom profiler. The inference loop terminates upon receiving a shutdown signal via a \texttt{threading.Event}.

This implementation achieves responsive and low-overhead inference while preserving data consistency. The adaptive frame-length mechanism complements the model's efficiency, allowing longer analysis windows only when acoustically justified.

\subsection{Full-Stack Live System with Remote Interface}

\begin{figure}
    \centering
    \includegraphics[width=1\linewidth]{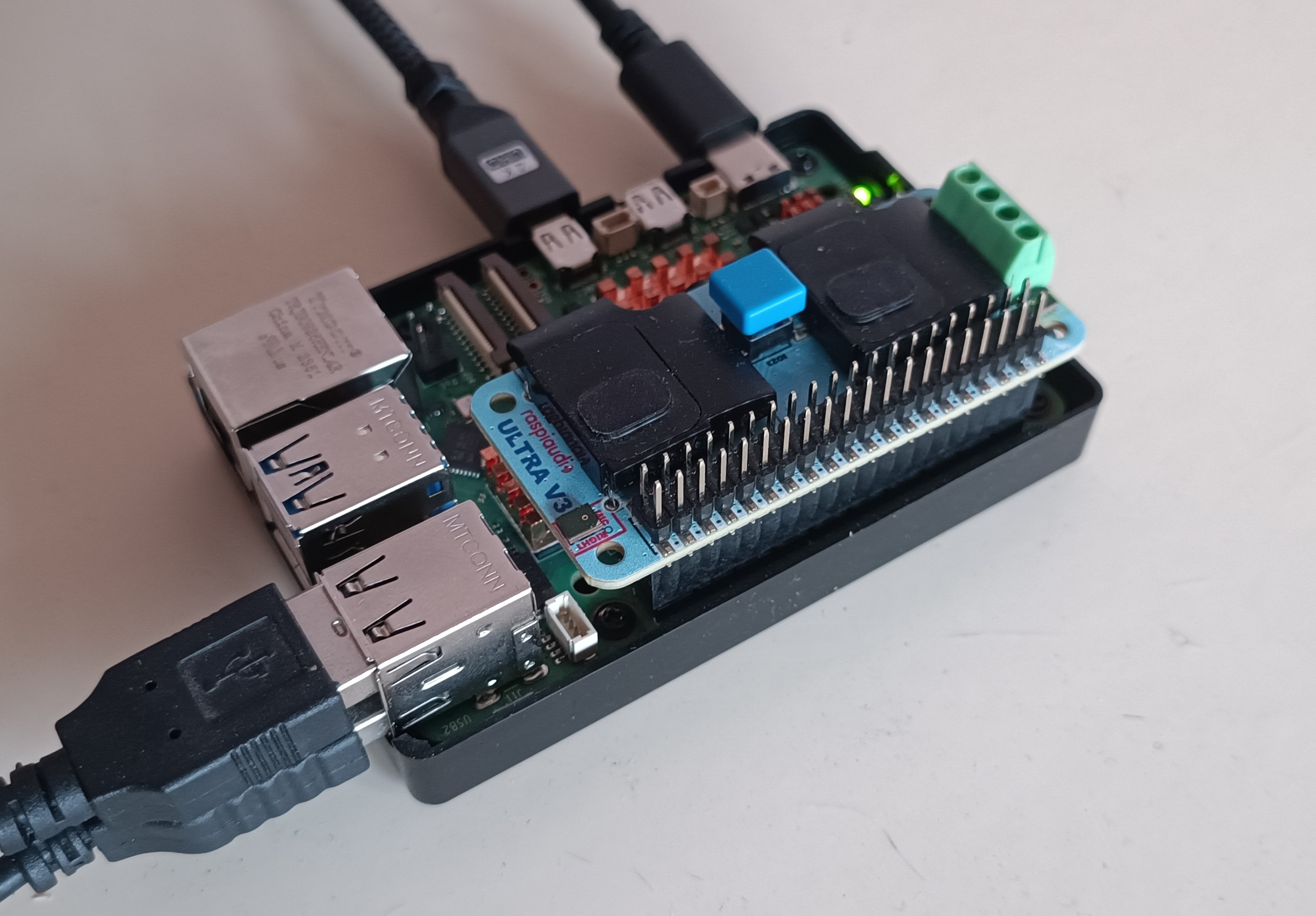}
    \caption{The Raspberry Pi 5 with RaspiAudio UltraV3 audio board on top}
    \label{fig:device}
\end{figure}

To transition from simulated real-time to a live operational setup, we integrated a RaspiAudio Ultra++ DAC+microphone audio board into the Raspberry Pi 5 platform. This board provides high-fidelity audio input with low-latency access to ALSA streams and onboard microphone array support (Fig. \ref{fig:device}).

\begin{figure}
    \centering
    \includegraphics[width=1.0\linewidth]{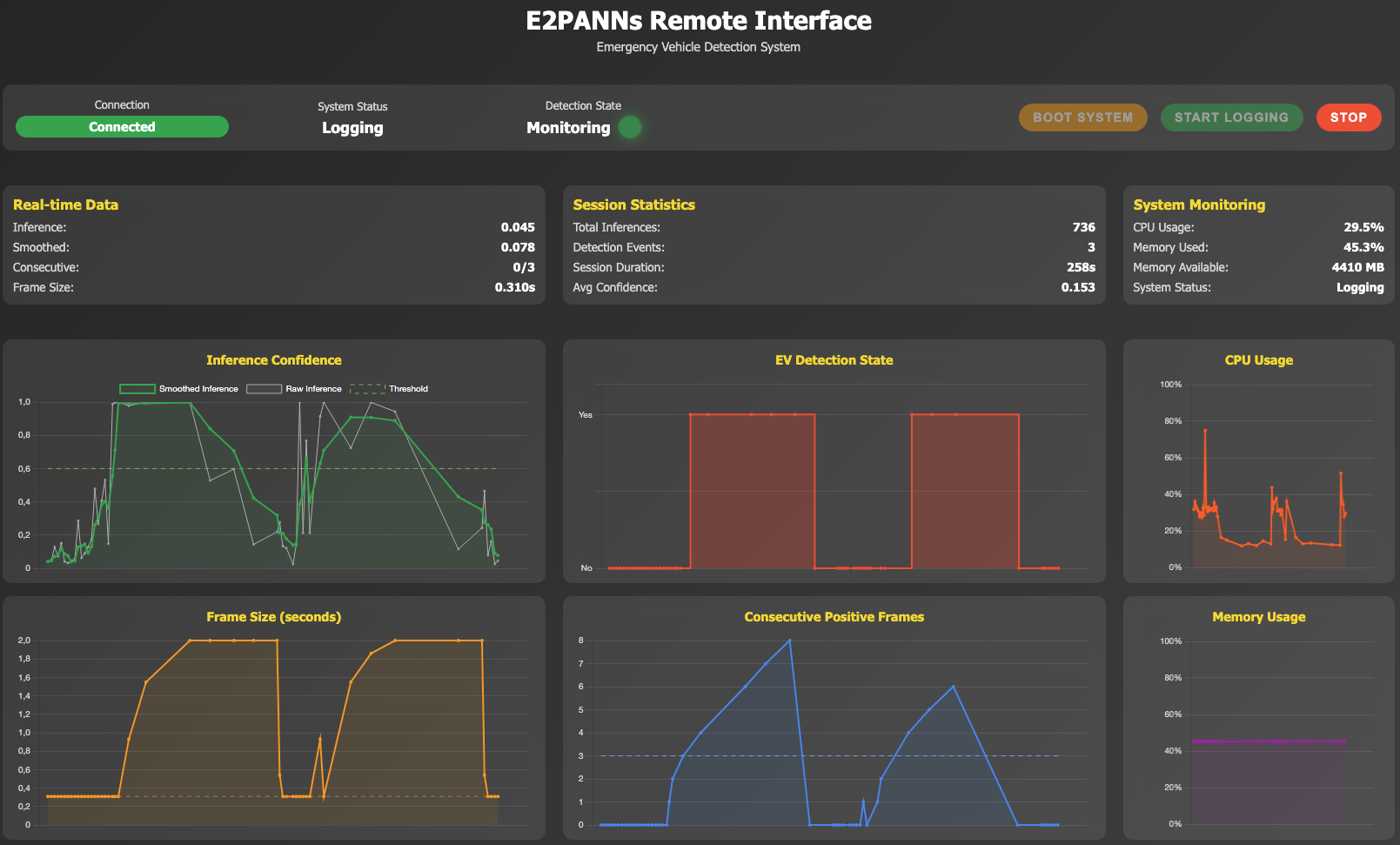}
    \caption{Layout of the graphic HTML/Javascript interface.}
    \label{fig:interface}
\end{figure}

The real-time engine is extended with a WebSocket-based interface for remote monitoring and control. The system exposes a live stream of classification probabilities, detection flags, and diagnostic metadata (e.g., current frame width, inference time per frame). The frontend, accessible via browser or terminal client, allows operators to inspect system behavior during field deployment or controlled playback scenarios (Fig. \ref{fig:interface}).

A core component of the postprocessing stage is the event decision state machine, which determines whether a positive siren detection event should be issued. This logic builds upon two key mechanisms that can be enabled and configured to meet specific needs:

\begin{itemize}
  \item \textbf{Moving average smoothing filter:} the sequence of raw inference probabilities is smoothed using a fixed-size moving average window to suppress transient noise and reduce the likelihood of spurious spikes triggering false positives.
  \item \textbf{Consecutive frame validation:} a detection event is confirmed only if a minimum number of consecutive frames (each exceeding a configurable probability threshold) are observed. This adds temporal consistency and reduces instability in event onset detection.
\end{itemize}

The state machine tracks ongoing classification outputs and triggers or resets detection flags accordingly, ensuring robust decision-making under noisy or ambiguous acoustic conditions. These design choices are critical for minimizing false alarms in edge deployment scenarios, particularly in urban environments.

This full-stack implementation supports live demonstrations, reproducibility experiments, and real-world data acquisition campaigns. It ensures all components — model, dataflow, post-processing, and human interaction — are co-located on a low-cost, power-efficient embedded platform.

Results from the real-time test scenarios and their SED performance analysis are detailed in section~\ref{sec:evaluation}.

\section{Evaluation of Real Time Sound Events Detection} \label{sec:evaluation}
\begin{figure}
    \centering
    \includegraphics[width=1\linewidth]{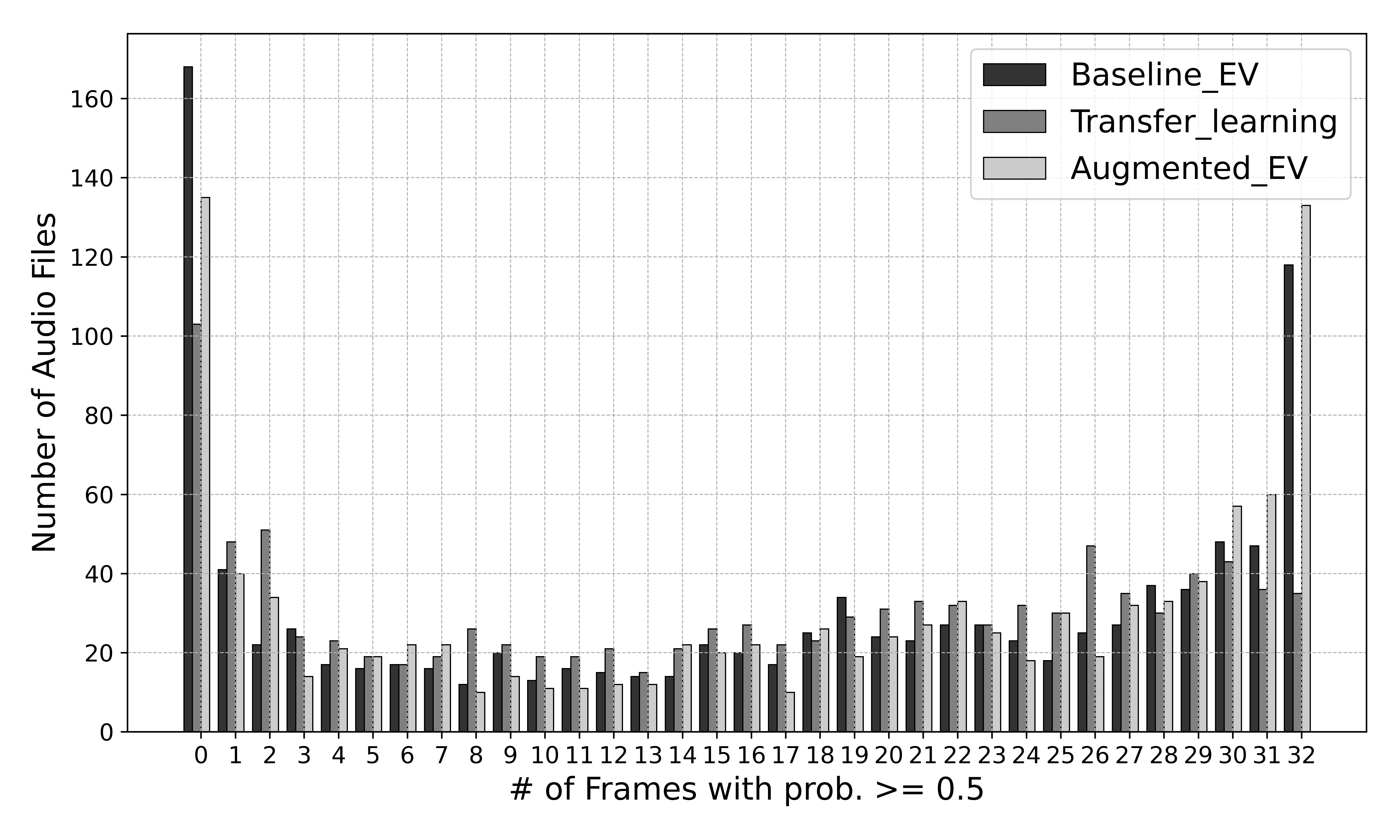}
    \caption{Distribution of samples based on the number of frames with probability $>=$ 0.5}
    \label{fig:frames_distribution}
\end{figure}

To assess the real-time performance of the E2PANNs model, we deployed it on a Raspberry Pi 5 8 GB (Broadcom BCM2712 2.4GHz quad-core 64-bit Arm Cortex-A76) within a custom multi-threaded inference pipeline, as seen in \ref{sec:edge-computing}. A ground truth evaluation was conducted using three checkpoints of the model (see \ref{subsec:e2panns}) and the 1,025 emergency vehicle (EV)-labeled samples from the AudioSet-Strong dataset \cite{Hershey2021TheBO}, a subset of AudioSet with temporally-strong labels, suitable for SED. Each sample underwent frame-wise simulated real-time inference with detailed logging of model outputs, frame durations, frame sizes, and system-level performance metrics including CPU load and memory usage. Three experimental conditions were evaluated across all three model checkpoints: one with fixed frame width, and two with variable frame width increasing at rates of 0.2 and 0.4 seconds per second of sustained over-threshold confidence predictions. A statistical analysis of the number of frames exceeding a 0.5 probability threshold (Figure \ref{fig:frames_distribution}) revealed a subset of 287 audio files with zero confident detections. These files were manually audited, and 182 (63.41\%) of them were confirmed to contain no emergency vehicle events despite their positive ground-truth labels. This human post-validation process exposed a significant presence of \textit{False} True Positives in the original dataset and led to the creation of a corrected subset, thereby enhancing the reliability of the test data and strengthening the credibility of subsequent performance evaluations. This finding underscores the importance of the availability of reliable datasets, particularly in safety-critical applications where false positives are more disruptive than missed detections.

To further assess the behavior of the E2PANNs model during real-time operation on the Raspberry Pi 5, we conducted a thorough evaluation of the inference outputs using standard Sound Event Detection (SED) metrics. The evaluation pipeline processed the frame-wise probability scores generated by the model on 1025 audio clips, each corresponding to a 10-second excerpt from YouTube videos containing emergency vehicle sounds.
To ensure compatibility with the input formats expected by the \texttt{torchmetrics} and \texttt{sed\_eval} libraries, the outputs were transformed into fixed-resolution binary sequences. This involved discretizing the fixed and variable-length inference windows (which depended on the adaptive frame duration logic) into uniform time steps and aligning them with the corresponding ground truth annotations. The annotations were drawn from a merged metadata file, which included manual validation of each event, and optionally filtered using a list of \textit{False} True Positives (i.e., predictions originally annotated as True Positives but subsequently determined to be incorrect). This allowed the evaluation to reflect a stricter reference standard when required.\\
Both frame-wise and event-based metrics were computed across all experiments and checkpoints. Frame-wise metrics — computed via \texttt{torchmetrics} \cite{detlefsen_torchmetrics_2022} — included precision, recall, F1-score, accuracy, specificity, balanced accuracy, and error rate. Event-based metrics — obtained using \texttt{sed\_eval} \cite{mesaros_sound_2021}— included event-level F-measure (and its precision and recall components) as well as error rate (and its insertion and deletion components). All metrics were aggregated and reported both at the individual checkpoint level and in a combined summary (Table \ref{tab:sed-evaluation-summary}) across the full evaluation set.\\

\begin{table*}[ht]
    \centering
    \caption{Evaluation summary for frame-wise and event-based metrics (EBM)}
    \label{tab:sed-evaluation-summary}
    \scriptsize
    \begin{threeparttable}
    \setlength{\tabcolsep}{3pt}
    \begin{tabular}{|c|c|c|c|c|c|c|c|c|c|c|c|c|c|c|}
    \hline
    \textbf{FTP} & \textbf{experiment} & \textbf{checkpoint} & 
    \multicolumn{6}{|c|}{\textbf{Frame-wise Metrics}} & 
    \multicolumn{3}{|c|}{\textbf{EBM f-measure}} & 
    \multicolumn{3}{|c|}{\textbf{EBM error rate}} \\
    \cline{4-15}
    & & & \textbf{precision} & \textbf{recall} & \textbf{f1} & \textbf{accuracy} & \textbf{specificity} & \textbf{bal\_accuracy} & 
    \textbf{f1} & \textbf{precision} & \textbf{recall} & 
    \textbf{error\_rate} & \textbf{deletion\_rate} & \textbf{insertion\_rate} \\
    \hline
True & const\_fr & (1) & \textbf{88\%} & 68\% & 72\% & 69\% & 24\% & 69\% & 14\% & 9\% & 26\% & 3.30 & 0.74 & 2.56 \\
True & const\_fr & (2) & 87\% & 64\% & 69\% & 65\% & 25\% & 65\% & 12\% & 8\% & 24\% & 3.36 & 0.76 & 2.60 \\
True & const\_fr & (3) & 86\% & 60\% & 66\% & 61\% & 26\% & 61\% & 5\% & 3\% & 12\% & 4.37 & 0.88 & 3.49 \\
True & var\_fr\_02 & (1) & 87\% & 74\% & 76\% & 73\% & 20\% & 73\% & 26\% & 21\% & 35\% & 2.00 & 0.65 & 1.35 \\
True & var\_fr\_02 & (2) & 87\% & 71\% & 73\% & 71\% & 21\% & 71\% & 24\% & 19\% & 32\% & 2.04 & 0.68 & 1.36 \\
True & var\_fr\_02 & (3) & 86\% & 61\% & 67\% & 62\% & 24\% & 62\% & 12\% & 9\% & 19\% & 2.82 & 0.81 & 2.01 \\
True & var\_fr\_04 & (1) & 87\% & \textbf{77\%} & \textbf{78\%} & \textbf{76\%} & 19\% & \textbf{76\%} & \textbf{30\%} & \textbf{25\%} & \textbf{38\%} & \textbf{1.74} & \textbf{0.62} & \textbf{1.12} \\
True & var\_fr\_04 & (2) & 87\% & 74\% & 75\% & 73\% & 19\% & 73\% & 27\% & 22\% & 34\% & 1.88 & 0.66 & 1.22 \\
True & var\_fr\_04 & (3) & 86\% & 63\% & 68\% & 64\% & 24\% & 64\% & 14\% & 10\% & 21\% & 2.60 & 0.79 & 1.81 \\
\hline
False & const\_fr & (1) & 76\% & 57\% & 60\% & 63\% & 31\% & 63\% & 13\% & 9\% & 22\% & 2.97 & 0.78 & 2.18 \\
False & const\_fr & (2) & 73\% & 53\% & 57\% & 60\% & \textbf{32\%} & 60\% & 12\% & 8\% & 20\% & 2.98 & 0.80 & 2.18 \\
False & const\_fr & (3) & 77\% & 50\% & 56\% & 57\% & \textbf{32\%} & 57\% & 5\% & 3\% & 10\% & 3.99 & 0.90 & 3.09 \\
False & var\_fr\_02 & (1) & 76\% & 62\% & 64\% & 67\% & 27\% & 67\% & 24\% & 20\% & 29\% & 1.89 & 0.71 & 1.18 \\
False & var\_fr\_02 & (2) & 73\% & 59\% & 61\% & 64\% & 28\% & 64\% & 22\% & 19\% & 27\% & 1.88 & 0.73 & 1.15 \\
False & var\_fr\_02 & (3) & 77\% & 52\% & 57\% & 58\% & 31\% & 58\% & 11\% & 8\% & 16\% & 2.65 & 0.84 & 1.81 \\
False & var\_fr\_04 & (1) & 76\% & 64\% & 66\% & 69\% & 26\% & 69\% & 27\% & 24\% & 32\% & 1.68 & 0.68 & 1.00 \\
False & var\_fr\_04 & (2) & 73\% & 61\% & 63\% & 66\% & 27\% & 66\% & 25\% & 22\% & 28\% & 1.75 & 0.72 & 1.04 \\
False & var\_fr\_04 & (3) & 77\% & 53\% & 58\% & 59\% & 30\% & 59\% & 12\% & 10\% & 17\% & 2.48 & 0.83 & 1.65 \\
    \hline
    \end{tabular} 
    \begin{tablenotes}
      \footnotesize
      \item Column \texttt{FTP} refers to checking against \textit{False} True Positives list.
      \item Checkpoint (1): audioset\_ev\_augmented; Checkpoint (2): audioset\_ev; Checkpoint (3): unified\_transfer\_learning.
      \item In bold the best result for each metric.
    \end{tablenotes}
    \end{threeparttable}
\end{table*}

To more rigorously evaluate the model’s resilience against spurious activations in real-world scenarios — where false alarms may lead to undesirable system behavior or user fatigue — we performed an in-depth false positive (FP) analysis grounded in the model’s inference frame structure, following the methodology proposed in previous large-scale SED benchmarks \cite{Mesaros_2018} \cite{Cakir_2017}. In this context, a false positive frame is defined as an output frame whose predicted probability exceeds the classification threshold but does not temporally overlap with any annotated ground truth event.

The computed metrics, summarized in Table \ref{tab:FP_analysis}, offer a multifaceted view of false positive behavior for each checkpoint over the three experiments. We report: (col. 3) the average number of false positive inference frames per sample (FP\_FW\_AFPS); (col. 4) the average proportion of these frames relative to the total number of inference frames (FP\_FW\_AFPSP); (col. 6) the total number of false positive events (FP\_EB\_T), defined as contiguous sequences of at least N consecutive FP inference frames (with $N=3$); (col. 7) the average model confidence associated with these FP events; (col. 8) the maximum observed run length of consecutive FP frames; and (col. 9) the mean run length over all events. Additionally, we computed (col. 5) the average model confidence over all inference frames (FP\_FW\_AC), providing a baseline against which the confidence of spurious activations can be contrasted.

\begin{figure}
    \centering
    \includegraphics[width=1\linewidth]{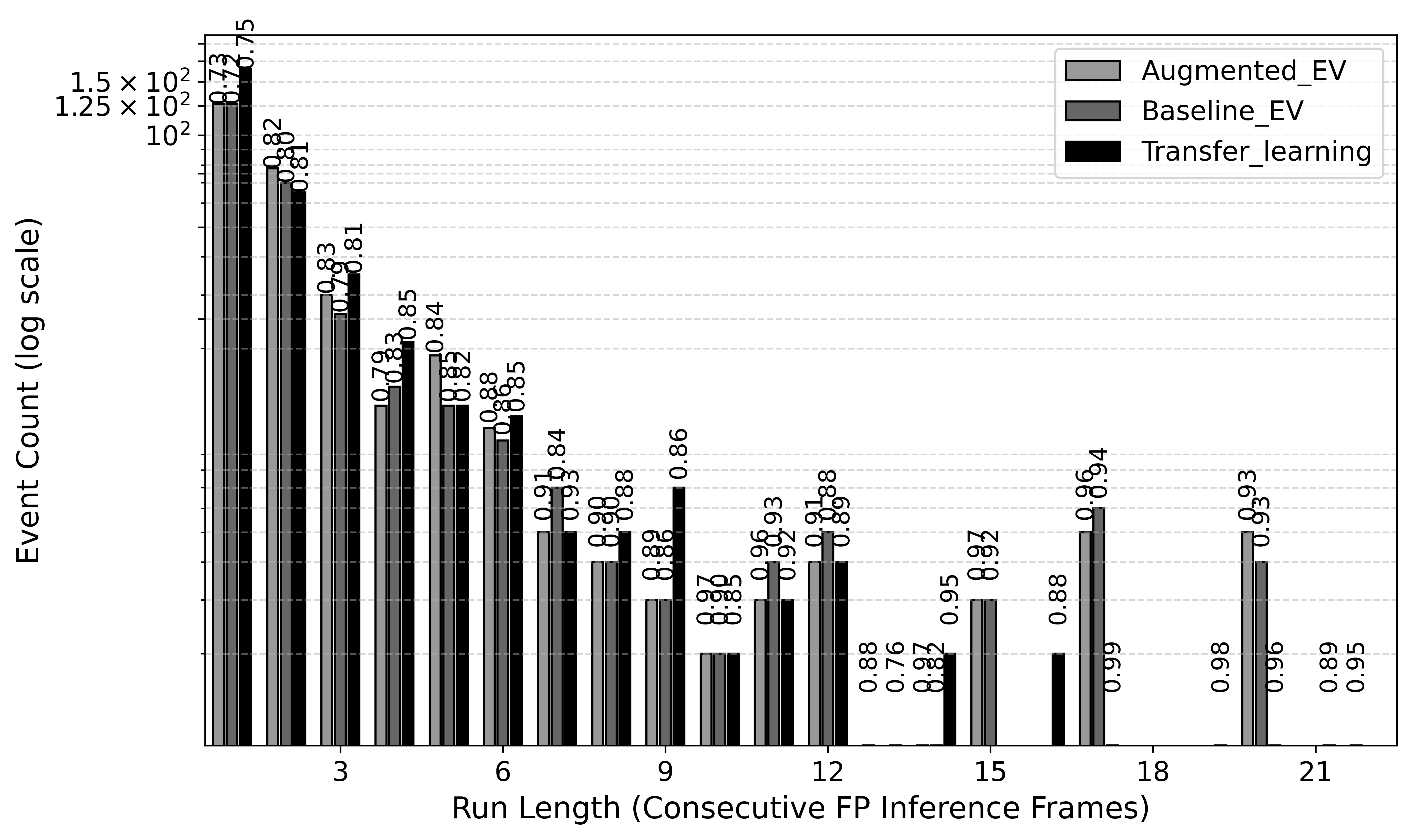}
    \caption{False Positives frequency distribution over events length for constant frame inference. Average confidence is reported above each bar.}
    \label{fig:fp_hist_constant_frame}
\end{figure}

\begin{figure}
    \centering
    \includegraphics[width=1\linewidth]{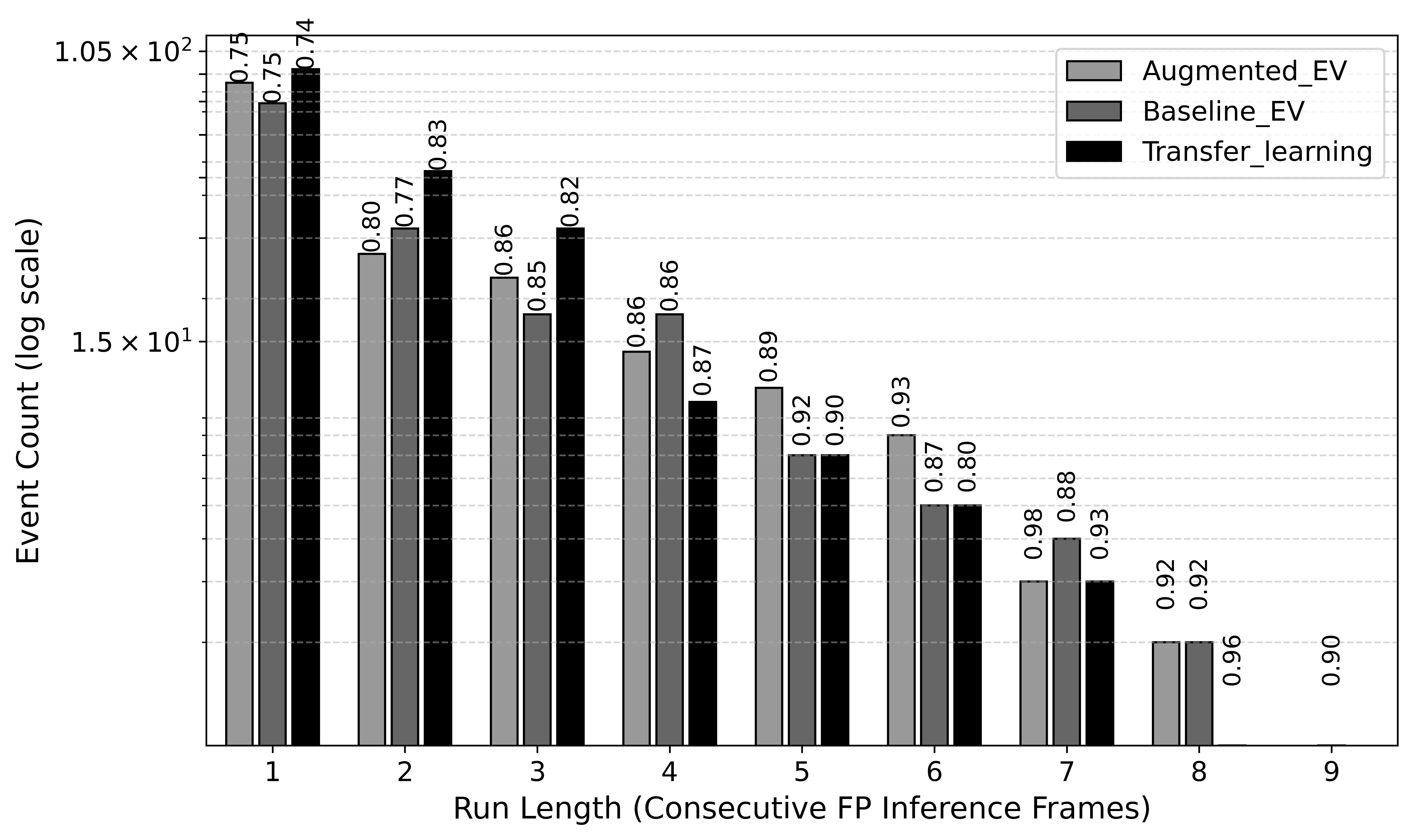}
    \caption{False Positives frequency distribution over events length for variable frame inference with \texttt{increment\_speed} $= 0.2$. Average confidence is reported above each bar.}
    \label{fig:fp_hist_var_frame_02}
\end{figure}

\begin{figure}
    \centering
    \includegraphics[width=1\linewidth]{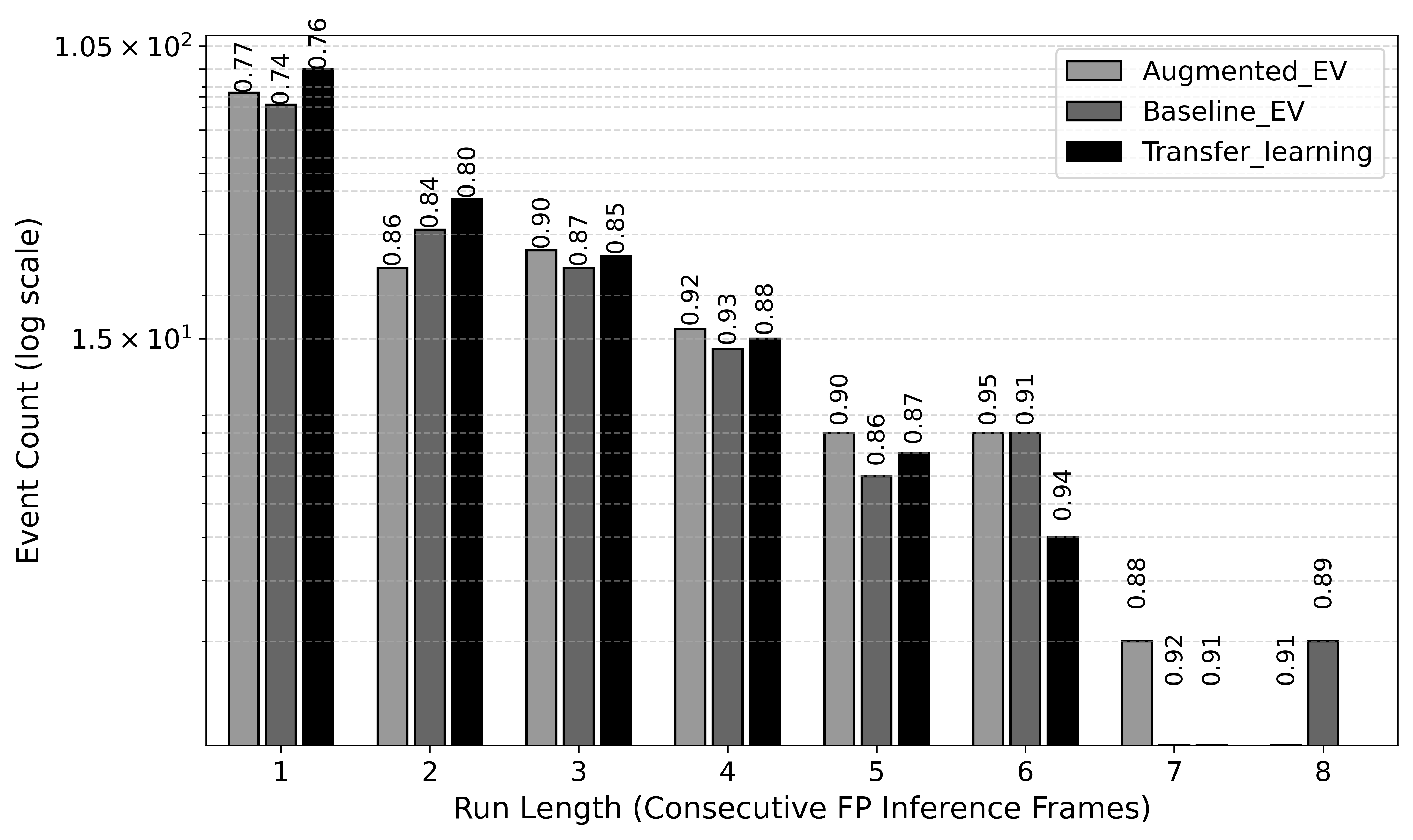}
    \caption{False Positives frequency distribution over events length for variable frame inference with \texttt{increment\_speed} $= 0.4$. Average confidence is reported above each bar.}
    \label{fig:fp_hist_var_frame_04}
\end{figure}

To complement these aggregate statistics, we generated per-experiment histograms \ref{fig:fp_hist_constant_frame} \ref{fig:fp_hist_var_frame_02} \ref{fig:fp_hist_var_frame_04} showing the frequency of FP event durations (in inference frames), stratified by checkpoint. These visualizations highlight the distribution and persistence of false positive bursts, with each bar annotated by the corresponding average model confidence. The resulting figures clearly indicate that the \texttt{constant\_frame} configuration is more prone to sustained and high-confidence spurious activations, whereas both \texttt{variable\_frame} configurations produce fewer and shorter false positive sequences. This supports the design hypothesis that adaptive windowing contributes to suppressing erroneous activations by leveraging temporally extended input segments, thus enhancing prediction stability in ambiguous regions.

\begin{table*}[ht]
    \centering
    \caption{False positives inference analysis}
    \label{tab:FP_analysis}
    \scriptsize
    \begin{threeparttable}
    \setlength{\tabcolsep}{3pt}
    \begin{tabular}{|c|c|ccc|cccc|}
        \hline
        \textbf{experiment} & \textbf{checkpoint} &
        \multicolumn{3}{c|}{\textbf{Frame-wise Based Metrics}} &
        \multicolumn{4}{c|}{\textbf{Event Based Metrics}} \\
        \cline{3-9}
        & & \textbf{FP\_FW\_AFPS} & \textbf{FP\_FW\_AFPSP} & \textbf{FP\_FW\_AC} &
        \textbf{FP\_EB\_T} & \textbf{FP\_EB\_AC} & \textbf{FP\_EB\_MRL} & \textbf{FP\_EB\_ARL} \\
        \hline
        const\_fr & (1) & 1.64 & 5.11\% & 68.47\% & 110 & 86.55\% & 22 & 7.06 \\
        const\_fr & (2) & 1.55 & 4.84\% & 65.17\% & 103 & 85.13\% & 20 & 7.15 \\
        const\_fr & (3) & 1.56 & 4.88\% & 58.73\% & 116 & 84.76\% & 21 & 6.16 \\
        var\_fr\_02 & (1) & 0.62 & 4.21\% & 62.69\% & 61 & 88.17\% & 8 & 4.34 \\
        var\_fr\_02 & (2) & 0.59 & \textbf{3.86\%} & 58.77\% & 55 & 87.06\% & 9 & 4.44 \\
        var\_fr\_02 & (3) & 0.64 & 3.94\% & \textbf{50.01\%} & 58 & \textbf{84.36\%} & 8 & 3.97 \\
        var\_fr\_04 & (1) & 0.59 & 4.45\% & 64.78\% & 62 & 90.92\% & 8 & 4.11 \\
        var\_fr\_04 & (2) & \textbf{0.56} & 4.11\% & 60.47\% & 55 & 89.26\% & 8 & 4.16 \\
        var\_fr\_04 & (3) & 0.57 & 3.97\% & 50.45\% & \textbf{53} & 86.75\% & \textbf{7} & \textbf{3.85} \\
        \hline
    \end{tabular}
    \begin{tablenotes}
      \footnotesize
      \item Checkpoint (1): audioset\_ev\_augmented; Checkpoint (2): audioset\_ev; Checkpoint (3): unified\_transfer\_learning
      \item FP\_FW\_AFPS: False Positives, frame-wise, average number of frames per sample
      \item FP\_FW\_AFPSP: False Positives, frame-wise, percentage of average number of frames per sample w.r.t. total frames
      \item FP\_FW\_AC: False Positives, frame-wise, average confidence
      \item FP\_EB\_T: False Positives, event based, total number of events
      \item FP\_EB\_AC: False Positives, event based, average confidence
      \item FP\_EB\_MRL: False Positives, event based, max run length
      \item FP\_EB\_ARL: False Positives, event based, average run length
    \end{tablenotes}
    \end{threeparttable}
\end{table*}

\begin{figure}
    \centering
    \includegraphics[width=1\linewidth]{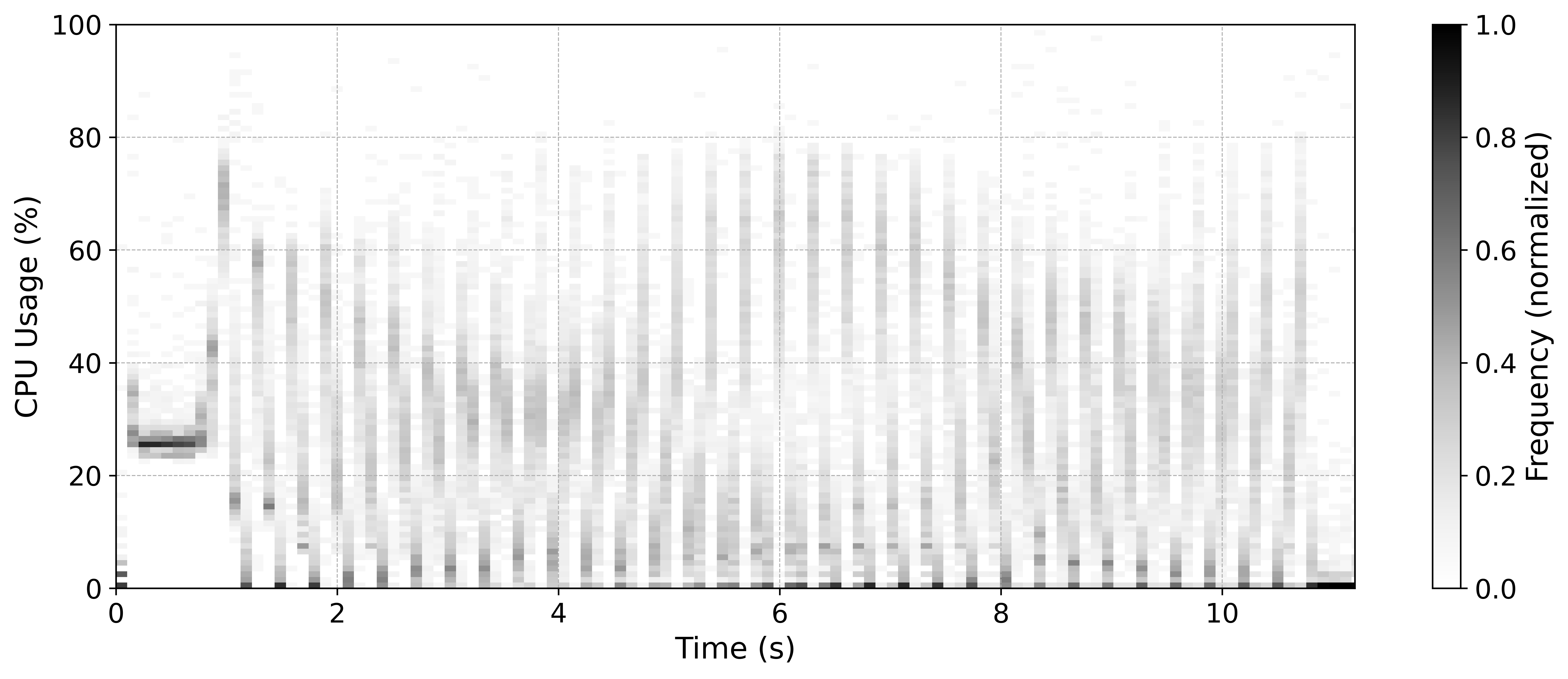}
    \caption{Heatmap of CPU load for constant frame experiment. Monitoring period $\sim 100ms$. Each frame displays the distribution of CPU load among test samples}
    \label{fig:cpu_heatmap}
\end{figure}

\begin{figure}
    \centering
    \includegraphics[width=1\linewidth]{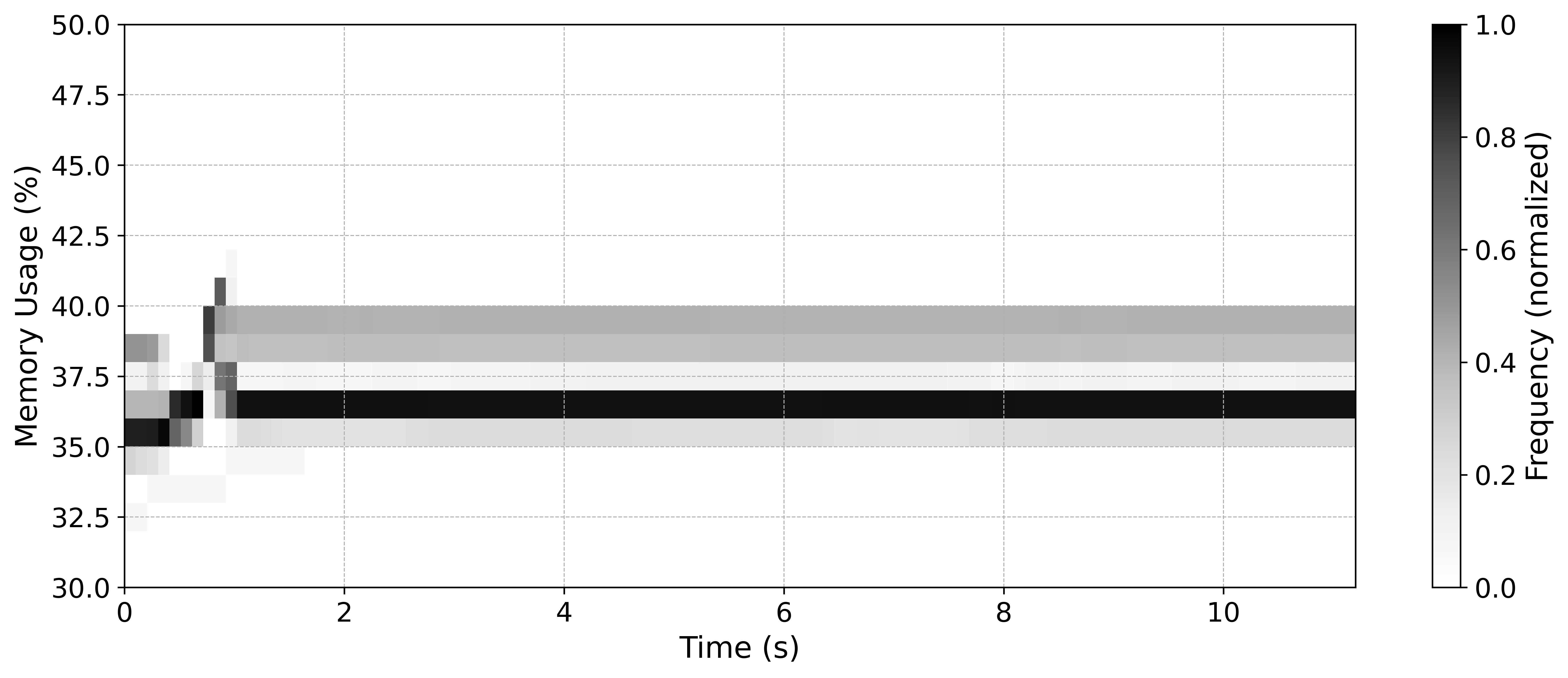}
    \caption{Heatmap of memory usage for constant frame experiment. Monitoring period $\sim 100ms$. Each frame displays the distribution of memory usage among test samples}
    \label{fig:memory_heatmap}
\end{figure}

Finally, to verify the feasibility of running the system in real time on the Raspberry Pi 5, we conducted a two-fold analysis. One one side we collected and analyzed CPU and memory load data during inference. From the corresponding log files, we generated heatmaps (Figures \ref{fig:cpu_heatmap}, \ref{fig:memory_heatmap}) showing the distribution of resource usage over time.
On the other side, to verify the feasibility of live deployment under operational conditions, we conducted two one-hour real-time experiments on the embedded system. The first was performed under an active soundscape containing siren and non-siren events (``multiple detections''), while the second was executed in an urban soundscape devoid of EV signals (``no detections''). The system statistics collected during both sessions are summarized in Table~\ref{tab:live-eval-summary}.

In terms of real-time processing performance, the system maintained a processing rate of 1.35x relative to frame duration under active detection, and a perfect 1.00x in the idle case. Maximum observed latency remained below 400 ms in both scenarios. The strong difference of \textit{total inferences} between the two experiments depends on the adaptive frame width mechanism. 

From the standpoint of stability, the system showed no runtime interruptions in either session. In the idle case, frame success was near perfect (99.53\%), with low CPU (30.3\%) and memory (15.5\%) usage. In contrast, the detection session demonstrated slightly higher load, but within acceptable thresholds.

No false detections were triggered during the baseline run (as expected, since no EV signals were present), while the active session reported 334 detection events, corresponding to a sustained detection rate of approximately 334 events/hour.

Taken together, these findings show that the system operates at the edge of real-time viability. Importantly, all measurements were obtained without applying low-level optimizations such as model quantization, ONNX conversion, or hardware-accelerated inference backends. These results confirm the viability of a full-stack SED pipeline on embedded ARM platforms and lay the foundation for further latency and efficiency improvements.

\begin{table*}[ht]
    \centering
    \makebox{
    \begin{threeparttable}
    \caption{Real-Time Performance and Detection Analysis: Comparative Evaluation}
    \label{tab:live-eval-summary}

    \scriptsize
    \renewcommand{\arraystretch}{1.1}
    \begin{tabular}{|llcc|}
    \hline
    \textbf{Category} & \textbf{Metric} & \textbf{Multiple Detections} & \textbf{No Detections} \\
    \hline
    \multicolumn{4}{|l|}{\textit{Real-Time Processing Performance}} \\
    
    & Normal Avg Frame Duration (ms) & 310.0 & 310.0 \\
    & Adaptive Avg Frame Duration (ms) & 848.5 & - \\
    & Avg. Processing Time (ms) & 318.1 & 309.9 \\
    & Overall Real-Time Factor & 1.35x & 1.00x \\
    & Normal Avg RT Factor & 1.00x & 1.00x \\
    & Adaptive Avg RT Factor & 2.48 & - \\
    & P95 Processing Time (ms) & 355.2 & 320.8 \\
    & P99 Processing Time (ms) & 370.6 & 328.3 \\
    & Max Processing Time (ms) & 442.7 & 370.0 \\
    \hline
    \multicolumn{4}{|l|}{\textit{System Stability Analysis}} \\
    & Session Duration (h) & 1.0 & 1.0 \\
    & Total Inferences & 7,249 & 11,553 \\
    & System Interruptions & 0 & 0 \\
    & Interruption Rate (per h) & 0.00 & 0.00 \\
    & Avg. Processing Rate (fps) & 2.0 & 3.2 \\
    \hline
    \multicolumn{4}{|l|}{\textit{Resource Usage Analysis}} \\
    & Avg. CPU Usage & 29.7\% & 30.3\% \\
    & Peak CPU Usage & 34.5\% & 37.3\% \\
    & Avg. Memory Usage & 16.0\% & 15.5\% \\
    & Peak Memory Usage & 16.5\% & 16.1\% \\
    \hline
    \multicolumn{4}{|l|}{\textit{Detection System Analysis}} \\
    & Total Detection Events & 334 & 0 \\
    & Detection Rate (events/h) & 333.9 & 0.0 \\
    \hline
    \end{tabular}
        \begin{tablenotes}
            \scriptsize
            \item \textbf{Real-Time Factor} is defined as \texttt{Frame\_Duration} / \texttt{Actual\_Processing\_Time}
            \item \textbf{P95 Processing Time} is the maximum duration of overall processing of 95\% of inference operations
            \item \textbf{P99 Processing Time} is the maximum duration of overall processing of 99\% of inference operations
        \end{tablenotes}
    \end{threeparttable}
    }
\end{table*}

\section{Live Demonstration Setup}

The system described in the previous sections has been deployed as a fully operational demonstrator, showcasing real-time emergency siren detection capabilities on a Raspberry Pi 5. The live demo environment replicates a realistic acoustic scenario by incorporating both audio playback of curated soundscapes and live microphone capture.

\subsection{Hardware Configuration}
The demonstration platform consists of:
\begin{itemize}
  \item A Raspberry Pi 5 (8 GB RAM) running a custom Linux-based OS image optimized for low-latency audio processing.
  \item A RaspiAudio Ultra++ DAC+mic board, providing high-quality microphone input and full ALSA compatibility (Fig. \ref{fig:device}).
  \item External speaker(s) for playback of controlled siren/non-siren audio samples.
\end{itemize}

The embedded model is executed in a multi-threaded inference pipeline (as detailed in Section \ref{sec:edge-computing}), enabling real-time classification on streamed audio without additional compute resources.

\subsection{Web-Based Remote Interface}
The system features a WebSocket-based remote interface for live monitoring and operator feedback. Built on a lightweight Python server backend, the interface exposes classification probabilities, detection triggers, and performance diagnostics (e.g., latency per frame, inference confidence trends).

The interface is accessible via browser and supports responsive updates every 100 ms, with visual elements including:
\begin{itemize}
  \item Real-time probability plot for the `Emergency Vehicle' class.
  \item Event detection flags based on state machine logic.
  \item Diagnostic display for current audio frame parameters (length, inference time).
\end{itemize}

This interface is essential for demonstration control and transparency, enabling reproducible SED testing, adjustable playback scenarios, and public engagement during the IEEE Internet of Sounds conference.

\section{Discussion, Conclusion, and Future Work}

\subsection{Label Reliability and Dataset Curation}
One of the most impactful bottlenecks encountered throughout the project was the poor reliability of the original tagging in Google's AudioSet corpus. Despite being one of the most comprehensive open-source audio datasets available, AudioSet suffers from weak labeling practices — many audio segments are annotated based on video-level metadata rather than precise acoustic content.

As a response, we introduced a curation pipeline using the AudioSet-Tools framework and developed the AudioSet-EV and AudioSet-EV Augmented datasets. These filtered and semantically structured distributions aim to mitigate label noise and enforce taxonomic clarity.

However, a problem became particularly evident during the evaluation of the AudioSet-Strong \cite{Hershey2021TheBO} subset, which was manually inspected to verify labeling quality. Numerous \textit{false} true positives and label inconsistencies were found in clips labeled as containing `Emergency Vehicle' sirens, significantly affecting both training and testing phases.

Taking all this into account, we believe that the broader issue of unreliable ground truth in large-scale audio corpora remains a fundamental limitation for training models in safety-critical domains.

\subsection{System Limitations and Deployment Challenges}
Another concern is the potential overfitting of the fine-tuned E2PANNs model to curated subsets. While evaluation on Unified-EV helped to assess generalization across diverse urban acoustic conditions, the performance may still degrade when exposed to highly variable environmental factors such as occlusions, reverberation, and background chatter. Further integration of data from public datasets and in-the-wild recordings may be necessary.

Although the embedded system performs within acceptable latency bounds, the need for multiple consecutive positive frames before declaring a detection introduces a trade-off between robustness and speed. In scenarios where siren cues are short-lived or partially occluded, delayed response could be critical.

Lastly, the effectiveness of detection is sensitive to the placement and quality of the microphone. In outdoor deployments or mobile installations, wind noise, vibration, and signal clipping can introduce errors. Future versions of the system could integrate beamforming or spatial filtering techniques.

\subsection{Summary of Contributions}
In this work, we presented a complete pipeline for emergency vehicle siren detection on embedded hardware, addressing both training and deployment challenges:
\begin{itemize}
  \item We evaluated and fine-tuned the EPANNs architecture for EV detection, obtaining the specialized E2PANNs model.
  \item We designed and released curated datasets (AudioSet-EV, AudioSet-EV Augmented, Unified-EV) using a reproducible, taxonomy-aware Python toolkit.
  \item We implemented a full real-time pipeline on Raspberry Pi 5 using multithreaded inference, adaptive frame logic, and post-processing filters.
  \item We evaluated the whole model on a temporally-strong labeled subset of AudioSet, finding, by human inspection, a relevant portion of \textit{false} true positives.
  \item We implemented a live system that integrates a high-quality DAC + mic board and a WebSocket remote interface for interactive monitoring.
\end{itemize}

\subsection{Future Work and Research Directions}
Several enhancements are currently under consideration:
\begin{itemize}
  \item Extension to multi-microphone setups enabling Direction-of-Arrival estimation and spatial filtering.
  \item Exploration of model distillation and quantization  techniques to further reduce inference latency.
  \item Deployment in moving vehicles and noisy environments to validate system resilience in the wild.
  \item Broader class coverage for urban sound events, moving beyond binary classification.
\end{itemize}

The IoS-ready architecture opens pathways for advanced distributed applications:
\begin{itemize}
    \item \textbf{Collaborative Detection Networks}: Multiple E2PANNs nodes can share acoustic fingerprints and detection confidence scores, enabling robust multi-point triangulation of emergency vehicles. Using edge computing federation, nodes can collaboratively reduce false positives through consensus mechanisms.
    \item \textbf{Smart City Integration}: The WebSocket interface enables direct integration with existing smart city platforms (FIWARE, Azure IoT Hub, AWS IoT Core), allowing emergency vehicle detections to trigger automated responses such as adaptive traffic signal control, public announcement systems, and emergency route clearance.
    \item \textbf{Acoustic Event Streaming}: By implementing Apache Kafka or similar event streaming platforms, our system can contribute to real-time acoustic maps of urban environments, where emergency vehicle movements are tracked alongside other sound events for comprehensive situational awareness.
    \item \textbf{Edge-Cloud Hybrid Processing}: While maintaining real-time edge detection, the system can selectively stream audio features to cloud services for advanced analytics, model updates, and long-term pattern analysis of emergency response times across the city.
\end{itemize}

These developments aim to support long-term goals of scalable, interpretable, and adaptive sound event detection in smart city infrastructures.

\bibliographystyle{IEEEtran}
\bibliography{bibliography}

\end{document}